\documentclass[aps,prb,twocolumn,superscriptaddress]{revtex4-2}
\usepackage[english]{babel}
\usepackage{bm}
\usepackage{amsmath}
\usepackage[dvips]{graphicx}
\usepackage[colorlinks=true, allcolors=blue]{hyperref}
\usepackage{diagbox}

\begin{document}
\title{Quasi-stationary near-gate plasmons in van der Waals heterostructures}
\author{A.~A.~Zabolotnykh}
\affiliation{Kotelnikov Institute of Radio-engineering and Electronics of the RAS, Mokhovaya 11-7, Moscow 125009, Russia}

\author{V.~V.~Enaldiev}
\affiliation{Kotelnikov Institute of Radio-engineering and Electronics of the RAS, Mokhovaya 11-7, Moscow 125009, Russia}
\affiliation{National Graphene Institute, University of Manchester, Booth St. E. Manchester M13 9PL, United Kingdom}
\affiliation{University of Manchester, School of Physics and Astronomy, Oxford Road, Manchester M13 9PL, United Kingdom}

\author{V.~A.~Volkov}
\affiliation{Kotelnikov Institute of Radio-engineering and Electronics of the RAS, Mokhovaya 11-7, Moscow 125009, Russia}

\begin{abstract}
Near-gate plasmons are a new type of plasma oscillations emerging in homogeneous two-dimensional electron systems where a gate provides partial screening of electron-electron interaction. Here we develop a theory of the near-gate plasmons in van der Waals heterostructures comprising  a conducting layer separated by a thin insulator from an uncharged disk-shaped gate. We show that in these structures the near-gate plasmons form gate-size-quantized quasi-stationary discrete modes even in the collisionless limit. Belonging to continuum spectrum of two-dimensional plasmons outside of the disk-gate, the near-gate plasmons are manifested as Fano-like resonances in frequency and magnetodispersions of scattering cross-section of the former scattered off the region under the gate. This enables to recover spectrum of the near-gate plasmons in the van der Waals heterostructures using near-field imaging techniques. 
\end{abstract}
\maketitle
\section{Introduction}

It is well known that properties of two-dimensional (2D) plasma oscillations (or plasmons) significantly depend on the characteristics of the medium surrounding 2D electron system. If 2D system is in dielectric environment then the plasmons possess a conventional square-root dispersion law~\cite{Stern1967}. However, for 2D system placed near the infinite metallic gate, the image charges, induced in the gate, screen strongly the Coulomb interaction between 2D electrons. As a result, the spectrum of such gated plasmons acquires ''acoustic'' behavior~\cite{chaplik1972}, with the frequency proportional to the 2D wave vector. 

Recently, it has been predicted a new type of plasmons -- near-gate plasmons -- emerging in \textit{partially} gated electron system, i.e. in an infinite homogeneous 2D system with a finite gate (for instance, in the form of a stripe or a disk), located nearby~\cite{zabolotnykh2019,zabolotnykh_disk_2019}.
In general, properties of the near-gate plasmons depend on the geometry of the gate. In the case of a stripe-shaped gate, spectra of these plasmons are determined by continuous longitudinal wave vector and discrete transversal number (starting from zero) equal to the number of nodes in charge density across the gate. The modes with non-zero transversal numbers have frequency gaps at zero longitudinal wave vector and they were investigated in Refs.~\cite{Iranzo2018, Satou2003,Satou2004,Ryzhii2006,Petrov2017,Davoyan2012,Bylinkin2019}. In contrast, the fundamental mode with zero transversal number is characterized by the specific gapless square-root dispersion~\cite{zabolotnykh2019,zabolotnykh2020} and it is the fundamental mode that is called the near-gate plasmon in Ref.~\cite{zabolotnykh2019}. However, in case of the gate confined in both lateral directions, we can expect that the differences, which existed between the fundamental and higher modes in the case infinite stripe-shaped gate when the longitudinal wave vector was continuous, are erased, since all plasmon modes become gate-size-quantized in both lateral directions. Therefore, in similar cases of confined geometry, for example, for the disk-shaped gate, we will refer to all modes as near-gate plasmons. Existence of such plasmons has been experimentally confirmed in 2D electron systems based on GaAs/AlGaAs quantum wells with stripe-shaped \cite{Muravev2019_2Dstripe,zarezin2020} and disk- or ring-shaped \cite{Muravev_ZarezinPRB2019,muravev2021crossover} geometry of the gate. Mention that these modes were called ''proximity plasmons'' in Refs.~\cite{Muravev2019_2Dstripe,zarezin2020,Muravev_ZarezinPRB2019,muravev2021crossover}.

However, a technique used to discover the near-gate plasmons in the conventional heterostructures \cite{Muravev2019_2Dstripe,zarezin2020,Muravev_ZarezinPRB2019,muravev2021crossover} enables to determine only their frequencies rather than spatial distribution of electric field for corresponding mode determined by quantization numbers. It may prevent extraction of the true dispersion of the near-gate plasmons, characterizing by orbital and radial quantization numbers, in the structures with disk-shaped gate, as the modes with the different numbers can have close eigen frequencies (as shown in Fig. \ref{fig:fig2}) and be indistinguishable for the experimental technique. 

\begin{figure}[t]
\includegraphics[width=1.0\columnwidth]{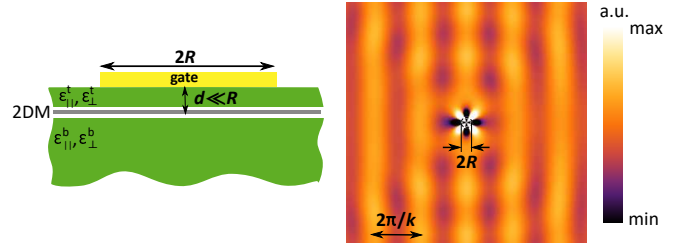}
    \caption{\label{fig:fig1} (Left panel) Side view of a vdW heterostructure (conducting 2D material encapsulated in dielectric environment) with disk-shaped metallic top-gate. Distance between gate and 2DM is assumed to be much smaller than the gate radius. (Right panel) Electric potential distribution for resonance scattering of 2D plasmons with the frequency corresponding to the quasi-stationary near-gate mode characterized by orbital momentum $l=4$.}
\end{figure}

The issue can be addressed in van der Waals (vdW) heterostructures \cite{geim2013,novoselov2016} with the help of local nano-imaging techniques, like scanning near-field microscopy \cite{chen2012,fei2012}, enabling a direct visualization of plasmon waves \cite{basov2016_Science,low2017polaritons}. In 2D material (2DM) plasmonics, graphene/hBN and hBN/graphene/hBN are the most studied vdW heterostructures \cite{woessner2015,grigorenko2012graphene,Basov2018} because of high carrier mobilities \cite{wang2013} and easy control of the spectral range of interest (from mid-infrared to visible) by means of gate voltage. However, multitude of two-dimensional materials discovered for the past decade, open up new possibilities \cite{song2021plasmons}.

Here, we demonstrate that the near-gate plasmons, forming in 2DM under an uncharged disk-shaped metallic gate (Fig. \ref{fig:fig1}), strongly modify an interference pattern, observable in the nano-imaging techniques \cite{basov2016_Science,low2017polaritons}, when the frequency of plasmon scattered off the region under the gate coincides with one of those of near-gate plasmons, as exemplified in Fig. \ref{fig:fig1}. In addition, we show that variation of a weak out-of-plane magnetic field can be used to tune the resonance conditions.

\section{Near-gate plasmons in zero magnetic field}

We consider a vdW heterostructure \footnote{Although the developed theory for near-gate plasmons is also applicable for 2D electron systems in the conventional heterostructures (like AlGaAs/GaAs), we believe that the predicted resonances in scattering cross-section is hard or even impossible to observe and visualize in these structures. This is because to provide high-mobility of carriers in the conventional structures, 2D conducting layer is placed deeply (on the order of 100 nm o more, see, for example, Refs.~\cite{muravev2021crossover, Shuvaev2021}) under the surface of the structures; while for the effective use of local excitation and measuring techniques the tip (for instance, in SNOM experiments) should be quite close to the 2D plane of free carriers (on the scale of 1-2 nm, see, for example, Ref.~\cite{zhao2021}).} 
consisting of a conducting 2DM thin film (graphene or mono(few)-layer transition metal dichalcogenides) encapsulated in dielectric environment and supplemented with a circular metallic gate of radius $R$, placed at a distance $d$ from 2DM (see Fig. \ref{fig:fig1}). Below, we study the near-gate plasmons using an approach used in Refs. \cite{Satou2003,Dyakonov2005,Petrov2016,Petrov2017,Jiang2018,rejaei2015,Dyer2012,Aizin2012,Dyer2013} where one finds self-consistent plasmon potential independently in several regions and then match the solutions with appropriate boundary conditions. In the quasi-static approximation, valid for long-wavelength plasmons far from the light cone, oscillating potential of plasmon propagating in 2DM, $\varphi e^{-i\omega t}$, is determined by the Poisson equation
\begin{equation}\label{Eq:Poisson}
    \left[\epsilon_{||}(z)\Delta_{r,\theta}+\partial_z\epsilon_{\perp}(z)\partial_z\right]\varphi(r,\theta,z)=-4\pi\rho(r,\theta)\delta(z),
\end{equation}
where $\epsilon_{||,\perp}(z)$ are in-plane and out-of-plane dielectric permittivities of insulators surrounding 2DM, $\epsilon_{||,\perp}(z)$ equals $\epsilon_{||,\perp}^{b}$ and $\epsilon_{||,\perp}^{t}$ at $z<0$ and $0<z<d$, correspondingly; $\Delta_{r,\theta}=\partial^2_r+(1/r)\partial_r+\partial^2_{\theta}$ is 2D Laplace operator in polar coordinates, $\rho(r,\theta)e^{-i\omega t}$ is 2D charge density of plasmon. The density is related with plasmon potential by the continuity equation in the layer plane $z=0$:
\begin{equation}\label{Eq:continuity_eq}
    i\omega\rho(r,\theta)+\sigma(\omega)\Delta_{r,\theta}\varphi(r,\theta,0)=0,
\end{equation}
where $\sigma(\omega)=iD/\pi\omega$ is a 2D local dynamic conductivity of 2DM expressed via the Drude weight $D$, which in the Drude model for electrons with effective mass $m^*$ is $D=\pi n_se^2/m^*$, while for massless Dirac electrons in graphene \cite{grigorenko2012graphene}, $D=e^2v\sqrt{\pi n_s}/\hbar$, where $n_s$ is 2D concentration of free charge carriers, $e$ is the elementary charge, and $v$ is the Dirac fermions' speed. Below we consider collisionless limit for conductivity $\omega\tau \gg 1$ neglecting scattering time $\tau$ in $\sigma(\omega)$. Combining \eqref{Eq:Poisson} and \eqref{Eq:continuity_eq} we obtain a single equation for the potential:
\begin{equation}\label{Eq:potential}
    \left[\epsilon_{||}(z)\Delta_{r,\theta}+\partial_z\epsilon_{\perp}(z)\partial_z\right]\varphi=\frac{4\pi\sigma(\omega)}{i\omega}\delta(z)\Delta_{r,\theta}\varphi.
\end{equation}

Under the disk-gate (at $r\leq R$), we require vanishing of plasmon potential at the gate surface 
\begin{equation}\label{Eq:gate_bc}
    \left.\varphi(r,\theta,d)\right|_{r\leq R}=0,
\end{equation}
assuming that gate is an ideal metal, i.e. it has infinitely large conductivity \footnote{For real metallic gates, having non-infinite conductivity, gate surface becomes equipotential on time scales determined by inverse 3D conductivity of the metallic gate (in CGS system having dimension of frequency). As long as the 3D conductivity is much higher than 2D plasmon frequency, the gate instantaneously screen the plasmon field, leading to zero boundary condition for the amplitude of plasmon potential at the gate surface. This issue was also considered in Discussion section of Ref.~\cite{zabolotnykh2019}}. The boundary condition \eqref{Eq:gate_bc} results in different wave vectors for a plasmon wave propagating in 2DM with the same frequency outside and under the gate, giving rise to elastic plasmon scattering. Note that the scattering mechanism occurs even for unbiased gate with zero total charge that does not introduce inhomogeneity in conductivity of 2DM.

Therefore, we look for solution of \eqref{Eq:potential} describing scattering of plasmon plane wave off the region under the gate. To this aim we solve \eqref{Eq:potential} separately in two regions (i) $r>R$ and (ii) $r\leq R$, and then match the two solutions at boundary $r=R$ in the plane of 2DM ($z=0$). 

Outside the 2DM plane plasmon potential satisfies the Laplace equation,
\begin{equation}\label{Eq:Laplace}
    \left[\epsilon_{||}^{t/b}\Delta_{r,\theta}+\epsilon_{\perp}^{t/b}\partial^2_z\right]\varphi=0,
\end{equation}
for which we impose a solution in a separable form,  $\varphi(r,\theta,z)=\varphi_{||}(r,\theta)\varphi_{\perp}(z)$. Substituting the latter in \eqref{Eq:Laplace} we obtain:
\begin{equation}\label{Eq:Laplace_separable}
    \frac{\Delta_{r,\theta}\varphi_{||}}{\varphi_{||}}+\frac{\epsilon_{\perp}^{t/b}}{\epsilon_{||}^{t/b}}\frac{\partial^2_z\varphi_{\perp}}{\varphi_{\perp}}=0.
\end{equation}
Introducing a separation parameter, $-k^2=\Delta_{r,\theta}\varphi_{||}/\varphi_{||}$, and, $k^2=\epsilon_{\perp}^{t/b}\partial^2_z\varphi_{\perp}/\epsilon_{||}^{t/b}\varphi_{\perp}$, we reduce \eqref{Eq:Laplace_separable} to the following system:
\begin{equation}\label{Eq:system}
\begin{cases}
    \left(\Delta_{r,\theta} + k^2\right)\varphi_{||}  = 0, \\
    \left(-\epsilon_{||}^{t,b}k^2+\epsilon_{\perp}^{t,b}\partial^2_z\right)\varphi_{\perp}=0, 
\end{cases}
\end{equation}
Here, the equation for $\varphi_{\perp}(z)$ should be supplemented by boundary conditions (BCs) at the plane $z=0$:
\begin{align}
\varphi_{\perp}(+0)&=\varphi_{\perp}(-0) \label{Eq:match1}\\
\epsilon_{\perp}(z)\partial_z \varphi_{\perp}(z)|^{z=+0}_{z=-0}& = -\frac{4\pi \sigma(\omega)k^2}{i\omega}\varphi_{\perp}(0). \label{Eq:match2}
\end{align}
where the first BC describes continuity of potential at the 2DM plane, and the second describes jump of electric field on the charged plane derived from \eqref{Eq:potential} by integration over $z$ with lower limit $-0$ and upper limit $+0$. Using Eq.~(\ref{Eq:system}) for $\varphi_{\perp}$ with BCs \eqref{Eq:match1} and \eqref{Eq:match2} we express the separation parameter $k$ via frequency outside and under the gate denoted $k(\omega)$ and  $k_{d}(\omega)$, respectively. Difference of the parameter in the two regions is due to an additional BC for $\varphi_{\perp}(z)$ at the gate surface $\varphi_{\perp}(d)=0$ (\ref{Eq:gate_bc}). 

Outside the gate ($r>R$), the in-plane part of the potential reads
\begin{equation}\label{Eq:in_potential_>R}
    \varphi_{||}(r,\theta) =\sum_{l=-\infty}^{+\infty}\left[i^lJ_l(kr)+i^lf_l(\omega)H_l^{(1)}(kr)\right]e^{il\theta},
\end{equation}
where $J_l(\xi)$ and $H_l^{(1)}(\xi)$ are respectively Bessel and the first kind Hankel functions, $f_l(\omega)$ are partial scattering amplitudes with orbital momenta $l=0,\pm1,\pm2,\dots$ Solution \eqref{Eq:in_potential_>R} corresponds to the boundary condition \cite{landau2013} of the scattering problem at large distances from the gate $kr\gg 1$: $$\varphi_{||}(r,\theta)= e^{ikr\cos\theta}+F(\omega,\theta)\frac{e^{ikr}}{\sqrt{ir}},$$ where the first term describes incident and passed plane wave, whereas the second term represents scattered wave characterized by the total scattering amplitude:
\begin{equation}\label{Eq:total_scattering_amplitude}
    F(\omega,\theta)=\sqrt{\frac{2}{\pi k(\omega)}}\sum_{l=-\infty}^{+\infty} f_l(\omega)e^{il\theta}.
\end{equation}

Using the above mentioned BCs~\eqref{Eq:match1} and \eqref{Eq:match2},
the out-of-plane potential at $r>R$ is expressed as
\begin{equation}\label{Eq:out_potential_>R}
\varphi_{\perp}(z)=
\left\{ 
\begin{split}
     &\varphi_0(0)\exp\left[kz\sqrt{\epsilon^b_{||}/\epsilon^b_{\perp}}\right], & z<0; \\
     &\varphi_0(z), & 0<z<d; \\
     &\exp\left[-k(z-d)\right], & z>d;
\end{split}
\right.
\end{equation}
where we choose normalization factor for plasmons potential such that $\varphi_{\perp}(d)=1$,
$$\varphi_0(z)=\cosh\left(k\sqrt{\frac{\epsilon^t_{||}}{\epsilon^t_{\perp}}}(z-d)\right) - \frac{\sinh\left(k\sqrt{\frac{\epsilon^t_{||}}{\epsilon^t_{\perp}}}(z-d)\right)}{\sqrt{\epsilon_{||}^t\epsilon_{\perp}^t}},$$ and $k(\omega)$ is determined by the following equation:
\begin{equation}\label{Eq:k_main}
    \sqrt{\epsilon_{||}^b\epsilon_{\perp}^b} + 
    \frac{\epsilon_{||}^t\epsilon_{\perp}^t \tanh\left( kd\sqrt{\frac{\epsilon_{||}^t}{\epsilon_{\perp}^t}}\right) +\sqrt{\epsilon_{||}^t\epsilon_{\perp}^t}} {\sqrt{\epsilon_{||}^t\epsilon_{\perp}^t}+\tanh\left( kd\sqrt{\frac{\epsilon_{||}^t}{\epsilon_{\perp}^t}}\right)} = \frac{4\pi \sigma(\omega)k}{i \omega}.
\end{equation}

In the long wave-length limit $kd\sqrt{\epsilon^t_{||}/\epsilon^t_{\perp}}\ll 1$ we obtain a standard relation between wave-number and frequency for the ungated 2D plasmon \cite{Stern1967}: 
\begin{equation}\label{Eq:k_limit}
    k=\frac{\omega^2}{4D}\left(1+\sqrt{\epsilon^b_{\perp}\epsilon^b_{||}}\right).
\end{equation}  

Under the gate, \mbox{$r<R$}, in-plane part of plasmon potential reads 
\begin{equation}\label{Eq:in_potential_<R}
    \varphi_{||}(r,\theta) =\sum_{l=-\infty}^{+\infty}A_lJ_l(k_dr)e^{il\theta},
\end{equation}
where coefficients $A_l$ are introduced to match $\varphi_{||}$ given by (\ref{Eq:in_potential_>R}) and (\ref{Eq:in_potential_<R}). Whereas the out-of-plane part of potential is 
\begin{multline}\label{Eq:out_potential_<R}
\varphi_{\perp}(z)= \\
\left\{ 
\begin{split}
     &-\sinh\left(k_dd\sqrt{\frac{\epsilon_{||}^t}{\epsilon_{\perp}^t}}\right)\exp\left(k_dz\sqrt{\frac{\epsilon^b_{||}}{\epsilon^b_{\perp}}}\right), & z<0; \\
     &\sinh\left(k_d(z-d)\sqrt{\frac{\epsilon_{||}^t}{\epsilon_{\perp}^t}}\right), & 0<z<d; \\
     & 0, & z>d;
\end{split}
\right.
\end{multline}
where $k_d(\omega)$ is determined by the equation \cite{chaplik1972}: 
\begin{equation}\label{Eq:k_d_main}
    \sqrt{\epsilon^b_{||}\epsilon_{\perp}^b}+\sqrt{\epsilon^t_{||}\epsilon_{\perp}^t}\coth\left(k_dd\sqrt{\frac{\epsilon_{||}^t}{\epsilon_{\perp}^t}}\right) = \frac{4\pi \sigma(\omega)k_d}{i \omega}.
\end{equation}

In the long wavelength limit $k_dd\sqrt{\epsilon^t_{||}/\epsilon^t_{\perp}}\ll 1$, (\ref{Eq:k_d_main}) describes acoustic 2D plasmons characterized by linear dispersion \cite{chaplik1972,alonso2017,lundeberg2017}:
\begin{equation}\label{Eq:k_d_limit}
    k_d=\frac{\omega}{V_p},
\end{equation}
with the velocity \mbox{$V_p=\sqrt{4Dd/\epsilon_{\perp}^t}$}.

As long as we consider 2DM with homogeneous conductivity we use simple conditions of continuity for in-plane potentials \eqref{Eq:in_potential_>R}, \eqref{Eq:in_potential_<R} and electric fields, \mbox{$\bm{E}_{||}=-\hat{r}\partial_r\varphi_{||}-\hat{\theta}\partial_{\theta}\varphi_{||}/r$} at the gate edge \mbox{$r=R$} in plane of 2DM, \mbox{$z=0$}. This gives
\begin{equation}
\label{Eq:scattering_amplitude}
 f_l(\omega)=\frac{J'_l(k_dR)J_l(kR)-\frac{k}{k_d}J'_l(kR)J_l(k_dR)}{\frac{k}{k_d}J_l(k_dR)H_l^{(1)'}(kR)-J_l'(k_dR)H_l^{(1)}(kR)},   
\end{equation}
\begin{equation}\label{Eq:amplitude}
    A_l=\frac{i^lJ_l(kR)+f_l(\omega)H_{l}^{(1)}(kR)}{J_l(k_dR)}.
\end{equation}

For a non-relativistic quantum particle poles of its scattering amplitude lying at negative energies determine energies of truly localized states in the system \cite{landau2013}. Although equation \eqref{Eq:system} for in-plane plasmon potential ($\varphi_{||}$) is formally identical to 2D Schr\"{o}dinger  equation, it describes only states with positive energies (as $k^2\propto \omega^4$ (\ref{Eq:k_limit}) outside gate and $k^2\propto \omega^2$ \eqref{Eq:k_d_limit} under the gate). Therefore, plasmons cannot form real localized modes under the gate but rather belong to continuum spectrum of plasmons outside the gate region. However, the continuum may host discrete quasi-stationary modes having complex frequencies $\omega'+i\omega''$ determined by complex poles of the scattering amplitude with $\omega'>0$ and $\omega''<0$ \cite{gamov1928,landau2013}. Axial symmetry of the problem allows us to characterize the quasi-stationary modes by orbital and radial quantization numbers. Equalizing denominator of $f_{l}(\omega)$ \eqref{Eq:scattering_amplitude} to zero we obtain a dispersion equation for these modes: 
\begin{equation}\label{Eq:disk_modes_dispersion}
    \frac{J'_l(k_dR)}{J_l(k_dR)}=\frac{k}{k_d}\frac{H_l^{(1)'}(kR)}{H_l^{(1)}(kR)}.   
\end{equation}
Right-hand side of \eqref{Eq:disk_modes_dispersion}, responsible for a decay of the near-gate plasmons, gives rise to quasi-stationary modes with large lifetimes ($|\omega''/\omega'|\ll1$) only for \mbox{$|k|R\ll 1$}. At \mbox{$|k|R\ll1$}  \eqref{Eq:disk_modes_dispersion} reduces to 
\begin{multline}\label{Eq:dispersion_eq_l_neq_0}
    \frac{J'_l(k_dR)}{J_l(k_dR)}\equiv\sum_{p=1}^{+\infty}\frac{2k_dR}{(k_dR)^2-\mu_{l,p}^2}+\frac{|l|}{k_dR}=\\
    =\left\{
    \begin{split}
    -\frac{|l|}{k_dR}+i\frac{2\pi \left(\frac{kR}{2}\right)^{2|l|}}{k_dR\,\Gamma^2(|l|)},&\quad|l|=1,2,3,\dots \\
    \frac{1}{k_dR\ln{\left(\frac{e^CkR}{2}\right)}-i\pi \frac{k_dR}{2}},&\quad l=0.
    \end{split}
    \right.
\end{multline}
where in the first identity we expressed logarithm derivative of the Bessel function  via $J_l$ zeros \cite{bateman1953}, $\mu_{l,p}$; $\Gamma(\xi)$ is the Gamma-function, and $C=0.577..$ is the Euler–Mascheroni constant. For every orbital momentum $l$, Eq.~\eqref{Eq:dispersion_eq_l_neq_0} has a series of roots enumerated by a radial quantization number $n=0,1,2,\dots$ in ascending order, characterizing the number of nodes at $r\leq R$ for potential of the $l$-th mode. The lowest quantized values for ${\rm Re}(k_d R)$ and, therefore, for the frequency~(\ref{Eq:k_d_limit}) of the near-gate quasi-stationary modes are listed in Table \ref{tab:quantization}. We note that allowed wave-numbers ${\rm Re}(k_d^{(n,0)})$ for axial symmetric modes have weak logarithmic dependence on wave-vector outside the gate, resembling spectra of the ''charged'' modes in 2D electron system electrically connected to gate \cite{Muravev2020physical}. 
Additional analysis of the obtained modes and the connection with the previous result~\cite{zabolotnykh_disk_2019} are given in Appendix\ref{AppNG}.

Since $k$ and $k_d$ are related with each other via the same frequency, lifetimes of the modes, determined by imaginary part in \eqref{Eq:dispersion_eq_l_neq_0}, strongly depend on magnitudes of quantization numbers themselves. Indeed, using equations for $k$ \eqref{Eq:k_limit} and $k_d$ \eqref{Eq:k_d_limit} in the long wavelength limit ($|k_d|d\sqrt{\epsilon_{||}/\epsilon_{\perp}}\ll1$), the condition for existence of long-lived near-gate plasmons is equivalent to  \mbox{$|kR|=|k_dR|^2(d/R)\left(1+\sqrt{\epsilon_{||}^b\epsilon^b_{\perp}}\right)/\epsilon_{\perp}^t\ll 1$}. The last strong inequality can be satisfied only for not too high quantum numbers in the limit \mbox{$d/R\ll 1$} (as \mbox{${\rm Re}(k_d)R\gtrsim 1$}, see Table \ref{tab:quantization}). Therefore, for real vdW heterostructures the total number of the near-gate plasmon modes is finite and determined by ratio $d/R$.  Inverse lifetimes of the near-gate modes can be estimated as follows:
\begin{equation}\label{Eq:plasmon_lifetimes}
    \omega''_{n,l} = 
    \left\{
    \begin{split}
    -\frac{\pi \omega'_{n,l}\left(\frac{{\rm Re}(k^{(n,l)})R}{2}\right)^{2|l|}}{2|l|\Gamma^2(|l|)}, & \, |l|=1,2,3,\dots\\
    -\frac{\pi \omega'_{n,0}}{4\left|\ln\left[\frac{e^C{\rm Re}(k^{(n,0)})R}{2}\right]\right|}, & \,l=0.
    \end{split}
    \right.
\end{equation}
Axial symmetric modes \mbox{$l=0$} possess the shortest lifetimes, while the lifetimes for modes with larger orbital momenta grow exponentially with $|l|$ as long as $|k^{(l,n)}|R\ll 1$. This is because modes with non-zero orbital momenta should tunnel through an additional barrier, produced by centrifugal motion, $(l^2-1/4)/r^2$, to decay into bulk continuum states outside of the gate.

\begin{table}
\centering
\caption{\label{tab:quantization} Quantization for wave vector and frequency ${\rm Re}(k_d^{(n,l)}R)=\omega'_{n,l} R/V_p$ of the quasi-stationary near-gate plasmon modes, computed from Eq.~(\ref{Eq:dispersion_eq_l_neq_0}) (values for axisymmetric modes, $l=0$, are given for $d/R=0.01$). }
\begin{tabular}{|c||c|c|c|c|c|}
\hline
\backslashbox{$n$}{$|l|$} & 0 & 1 & 2 & 3 & 4\\
\hline
0 & 0.59\, & 2.43\, & 3.87\, & 5.18\, & 6.44\, \\
1 & 3.95\, & 5.57\, & 7.08\, & 8.49\, & 9.85\, \\
2 & 7.10\, & 8.74\, & 10.27\, & 11.73\, &  13.13\, \\
\hline
\end{tabular}
\end{table}

\begin{figure}
\includegraphics[width=1.0\columnwidth]{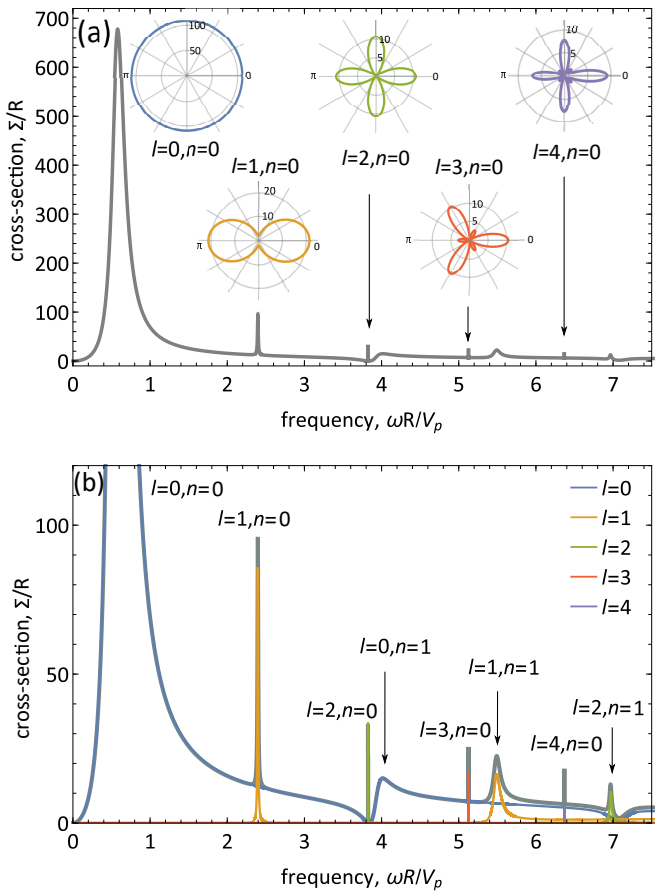}
\caption{\label{fig:fig2} (a) Frequency dependence of total scattering cross-section (\ref{Eq:scattering_cross_section}) for $d/R=0.01$. Resonances occur at frequencies of quasi-stationary plasmon modes listed in Table \ref{tab:quantization}. Insets show angle distribution for differential cross-section of the lowest modes for the first five orbital momenta. (b) Zoom-in of (a) with resonances indexed by orbital and radial quantization numbers of the quasi-stationary near-gate plasmon modes. Different colors show contributions of partial scattering cross-sections for indicated orbital momenta. For calculation we used dielectric permittivities of hBN, $\epsilon^{t,b}_{||}=6.9$, $\epsilon^{t,b}_{\perp}=3.6$ \cite{laturia2018}.}
\end{figure}

Having established resonance features of the scattering amplitude, we analyze a total scattering cross-section, $\Sigma$, characterizing ratio of plasmon energy densities of outgoing cylindrical and incident plane waves integrated over far distant circle (with radius $r\gg 1/k$) comprising the gate: 
\begin{multline}\label{Eq:scattering_cross_section}
\Sigma(\omega) = \int_{0}^{2\pi}rd\theta\frac{k^2\left|F(\omega,\theta)\right|^2}{rk^2}=\int_0^{2\pi}\left|F(\omega,\theta)\right|^2d\theta = \\
=\frac{4}{k}\left[\left|f_0\right|^2+2\sum_{l=1}^{+\infty}\left|f_l\right|^2\right],
\end{multline}
where the factor of two in brackets after the last equality results from relation, $f_l=f_{-l}$, responsible for double degeneracy of the near-gate modes with non-zero orbital momenta. For the case of long-lived near-gate plasmons, \mbox{$kR\ll1$}, partial scattering cross-sections takes an asymmetric form of Fano-resonance \cite{fano1935,fano1961}:
\begin{equation}\label{Eq:Fano_resonance}
    \frac{1}{k(\omega)}|f_l(\omega)|^2=
    \frac{1}{k(\omega)}\frac{\left[\chi_l(\omega) X_l(\omega)-1\right]^2}{\frac{X_l^2(\omega)}{\gamma_l^2(\omega)}+1},
\end{equation}
where function
\begin{multline}
   X_l(\omega)=\sum_{p=1}^{+\infty}\frac{2k_d(\omega)R}{(k_d(\omega)R)^2-\mu_{l,p}^2}+\frac{2|l|}{k_d(\omega)R} + \\
   +\frac{\delta_{l,0}}{k_d(\omega)R\left|\ln\left(\frac{e^Ck(\omega)R}{2}\right)\right|}
\end{multline} 
has plain zeros at frequencies of the near-gate modes $\omega=\omega'_{n,l}$ (see \eqref{Eq:dispersion_eq_l_neq_0}), and
\begin{equation}
    \gamma_{l}=\frac{2\pi \left(\frac{kR}{2}\right)^{2|l|}}{\Gamma^2(|l|)k_dR}\left(1-\delta_{l,0}\right) + \frac{\pi\delta_{l,0}}{2k_dR\left[\ln\left(\frac{e^CkR}{2}\right)\right]^2}
\end{equation}
determines the resonance width, which is directly proportional to lifetimes of the near-gate plasmons \eqref{Eq:plasmon_lifetimes}; the function $\chi_l=k_dR\left[(1-\delta_{l,0})/2|l|+\delta_{l,0}\left|\ln\left(e^CkR/2\right)\right|\right]$ is introduced for convenience. 
Fano resonances in plasmon scattering arise due to the interference of potential scattering between states of continuum with resonance scattering of continuum states on quasi-discrete states under the disk-shape gate. Such interference does not emerge for the stripe-shaped gate systems as for this geometry of the structure component of plasmon wave-vector along the stripe is a good quantum number. This leads to possibility of only 1D scattering with a single passed/reflected wave and, therefore, absence of Fano resonances in transmission and reflection coefficients.
Close to resonances $\omega\approx\omega'_{n,l}$ the total scattering cross-section, shown in Fig. \ref{fig:fig2}, is dominated by a corresponding partial contribution \eqref{Eq:Fano_resonance}. Therefore, peak magnitude of the total cross-section decays with frequency as $1/k=1/\omega^{2}$ (as $|f_l(\omega'_{n,l})|^2=1$), and can be used to extract $n$ of the near-gate plasmon mode for given $l$. Whereas, angle distribution of the differential cross-section, $|F|^2$, shown in insets of Fig. \ref{fig:fig2}, allows one to characterize its orbital momentum. At resonances $\omega=\omega_{n,l}'$, angular distributions of the total scattering amplitude have in-plane symmetry of the $l$-th cosine harmonics, as they are mainly determined by  $\left|f_0(\omega_{n,l}')+2f_l(\omega_{n,l}')\cos(l\theta)\right|^2$. However, this symmetry may be broken by an admixture of other small partial amplitudes, as it happens for the resonance \mbox{$l=4,n=0$} where non-vanishing contribution of $f_1(\omega_{0,4}')$ (see Fig. \ref{fig:fig2}(b)) results in asymmetry for scattering on 0- and $\pi$-angles. 

Thus, to recover spectrum of the near-gate plasmons in disk-shaped geometry one can combine frequency spectroscopy with the angular distribution analysis of scattering. 

\section{Effect of a weak out-of-plane magnetic field}

In a weak out-of-plane magnetic field $\sigma(\omega)$ in \eqref{Eq:continuity_eq} should be replaced with \cite{Crassee2011}
\begin{equation}
\label{Drude_xx}
    \sigma_{xx}(\omega)=\frac{D}{\pi}\cdot\frac{i\omega}{\omega^2-\omega_c^2},
\end{equation}
where $\omega_c=|e|B /(m_c c)$~\cite{Ando2002} is the electron cyclotron frequency ($m_c=\hbar\sqrt{n_s\pi}/v$ is the cyclotron mass for Dirac fermions in graphene, and $m_c=m^*$ for massive electrons). The off-diagonal (Hall) component of 2DM conductivity does not contribute in the continuity equation~\eqref{Eq:continuity_eq}, making equation for the plasmon potential~\eqref{Eq:potential} independent on orientation $\pm |B|$ of magnetic field.   

The scattering amplitude and cross-section are determined by \eqref{Eq:total_scattering_amplitude}, \eqref{Eq:scattering_cross_section} where in definition of wave-numbers $k$ \eqref{Eq:k_limit} and $k_d$ \eqref{Eq:k_d_limit} one should substitute $\omega\to\sqrt{\omega^2-\omega_c^2}$, which corresponds to $\sigma(\omega)\to\sigma_{xx}(\omega)$. 

\begin{figure}[t]
\includegraphics[width=1.0\columnwidth]{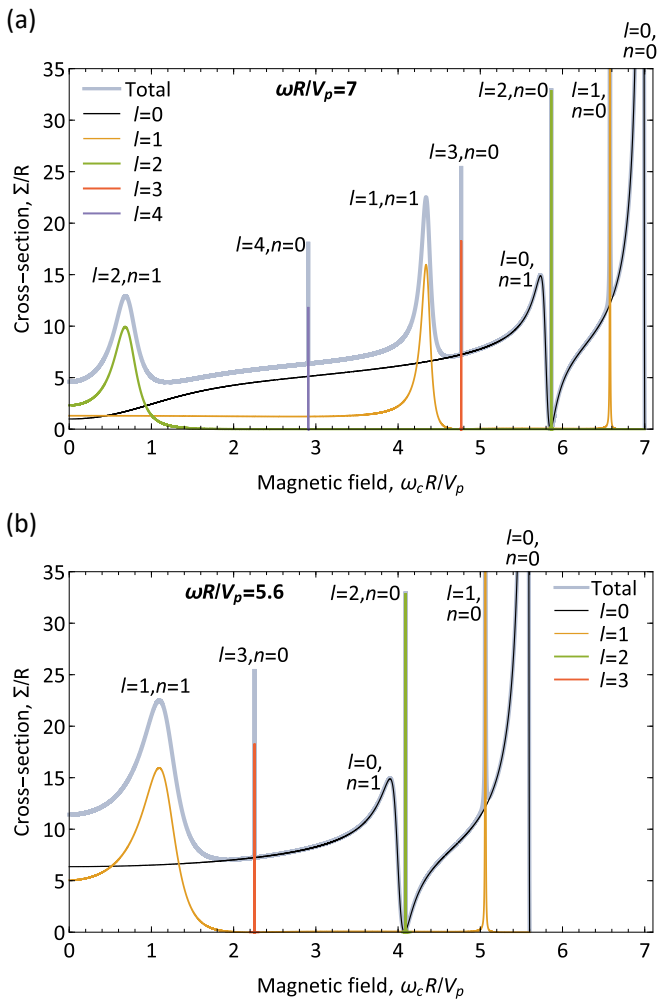}
    \caption{\label{fig:fig3} Magnetodispersion of total scattering cross-section (solid gray line) and contributions from partial cross-sections (\ref{Eq:scattering_cross_section}) with $l=0,1,2,3,4$ shown by different color. Figs. (a) and (b) are plotted for frequencies of the excited plasmon wave $\omega R/V_p$ equal to $7$ and $5.6$, correspondingly; other parameters of the system are the same as in Fig.~\ref{fig:fig2}.
    Resonances are indexed by orbital $l$ and radial $n$ quantization numbers of the quasi-stationary near-gate plasmon modes.
    All resonances have magnitudes as corresponding resonances at zero magnetic field, see Fig.~\ref{fig:fig2}.
     }
\end{figure}

Magnetodispersion of the total scattering cross-section, shown in the Fig. \ref{fig:fig3}, possesses resonances at cyclotron frequencies
\begin{equation}\label{Eq:omegac_res}
    \omega_c^{n,l}=\sqrt{\omega^2-(\omega'_{n,l})^2},
\end{equation}
where $\omega$ is the frequency of incident plasma wave and $\omega'_{n,l}$ is the resonance frequency of the corresponding near-gate plasmon mode in zero magnetic field. The number of resonances in the magnetodispersion is determined by magnitude of excitation frequency as that for freely propagating plasmons should always exceed cyclotron frequency $\omega_c$, providing real values of $k$. Equation \eqref{Eq:omegac_res} can be alternatively interpreted as a shift of the cross-section resonances in Fig. \ref{fig:fig2} induced by magnetic field. This allows one to control resonance conditions varying the field magnitude. Since the shift does not modify peak values of the resonances for the same $n$ and $l$, magnetodispersion  of the total cross-section can be also used to extract dispersion of the near-gate plasmon modes similarly as it was discussed above.

It is interesting to note that resonance widths in magnetodispersion of the plasmon cross-section can be tuned by the choice of the excitation frequency. This is because dispersion of gated plasmons in magnetic field, $\omega^2=\omega_c^2+V_p^2k_d^2$, leads to wave-number dependent group velocity. Therefore, the width of a resonance (with fixed numbers $n$ and $l$) in magnetodispersion of the cross-section increases with the decrease of frequency $\omega$.
In Fig.~\ref{fig:fig3} we demonstrate this behaviour comparing magnetodispersions of the scattering cross-section for two different excitation frequencies. 
  
Although in out-of-plane magnetic field the eigen frequencies of near-gate plasmons gain a conventional blue shift because of cyclotron motion, $\omega_{n,l}(B \neq 0)=\sqrt{(\omega_{n,l}')^2+\omega_c^2}$ (\ref{Eq:omegac_res}), it is useful to compare their magnetic field dependence with plasmons in a fully gated 2D system in the form of a disk, where plasmon modes are also characterized by orbital and radial quantization numbers.

In magnetic field $\bm{B}$ the frequencies of plasmons in disk-shaped 2DM are split into two branches due to the sign of the orbital number $l$, see Fig. 1 in Ref.~\cite{Fetter1986}, which leads to a rather complicated structure of plasma modes, including the appearance of so-called edge magnetoplasmons Refs.~\cite{Mast1985,Glattli1985,Volkov1985,Volkov1988}. On the contrary, for near-gate plasmons there is no splitting in magnetic field due to the sign of $l$, which makes the application of magnetic field a convenient tool for studying near-gate plasmons.

\section{Discussion and conclusion}

Existence of the discrete near-gate plasmon modes under the disk-gate at zero gate voltage, i.e. for homogeneous 2DM, makes these structures attractive for applications, because the resonance frequencies can be readily tuned by a size of gate. While lack of translational-invariance in the whole structure relaxes momentum conservation rules allowing direct excitation of plasmons by photons independently on their polarisation. It is likely that the near-gate plasmons have been recently observed in graphene nanopatterned with array of metallic nanocubes \cite{Koppens2020} (see also~\cite{Kaydashev2021}).

In our analysis we assume collisionless limit for 2DM conductivity $\omega\tau\to\infty$, whereas finite electron momentum relaxation time $\tau$ would reduce lifetimes \eqref{Eq:plasmon_lifetimes} of the near-gate modes. So, it is interesting to compare the lifetimes with typical values of electron relaxation time $\tau$ in 2DM. Lifetimes (\ref{Eq:plasmon_lifetimes}) for the \mbox{$n=0$}, \mbox{$l=0$}, and \mbox{$n=0$}, \mbox{$l=1$} modes are \mbox{$\left(\omega''_{00}\right)^{-1}=R/(0.086V_p)$} and \mbox{$\left(\omega''_{01}\right)^{-1}=R/(0.009V_p)$}, respectively. Considering graphene as 2DM and taking parameters close to those of Ref.~\cite{Koppens2020}, $d=2$\,nm, $E_F=0.47$\,eV, and $R=50$\,nm, we find frequencies $\omega'_{00}/2\pi=3.4$\,THz, $\omega'_{01}/2\pi=14$\,THz and lifetimes $\left(\omega''_{00}\right)^{-1}=0.31$\,ps and $\left(\omega''_{01}\right)^{-1}=3$\,ps. These lifetimes substantially exceed relaxation time $\tau=10$\,fs (at $T=300$\,K) in the system \cite{Koppens2020}, making electron collisions a dominant factor of the near-gate mode decay. However, in principle the electron relaxation time in graphene can be of the order of $1$\,ps \cite{Bolotin2008,Dean2010}. Consequently, in this case decay rate of the near-gate plasmons is due to their quasi-stationary feature.

We also mention that the used and similar approaches \cite{Satou2003,Dyakonov2005,Petrov2016,Petrov2017,Jiang2018,rejaei2015,Dyer2012,Aizin2012,Dyer2013}, namely, solving Eq.~(\ref{Eq:system}) for $\varphi_{||}$ in separate regions with the subsequent matching at the boundary, are widely exploited to study plasmons in 2D systems with inhomogeneous profile of concentration (or conductivity), despite of the fact that they lead to discontinuity of potential outside of the conducting plane, due to difference in the out-of-plane behaviour of plasmon potential in the regions with different concentrations. Although in our system charge density in 2DM is homogeneous, screening of plasmon potential under the disk-gate gives rise to the same discontinuity at the gate edge outside of the 2DM plane. Nevertheless, the obtained expressions for scattering amplitude \eqref{Eq:scattering_amplitude} and cross-section \eqref{Eq:scattering_cross_section} are independent on the choice of the matching plane (below the disk-gate), and lead to the same eigen frequencies for the axial asymmetric (i.e. $l\neq 0$) quasi-stationary near-gate modes as were obtained in a different approach \cite{zabolotnykh_disk_2019}. 

To conclude, we demonstrated that uncharged metallic gate, providing partial screening of electron-electron interaction in homogeneous conducting 2DM, gives rise to discrete near-gate plasmon modes, possessing finite lifetimes, and, for the disk-shaped gate, quantized by its radius.  Characterized by radial and orbital quantization numbers, the near-gate modes are manifested as Fano-like resonances in cross-section of 2D plasmons scattered off the region under the gate. Out-of-plane magnetic field controls the resonance frequencies allowing to tune their widths in magnetodispersion of the scattering cross-section. 

\begin{acknowledgments}
    We are grateful to Igor Zagorodnev for valuable discussions.
    The work was financially supported by the Russian Science Foundation (Project No. 21-12-00287).
\end{acknowledgments}

\appendix*
\section{Near-gate plasmons in different 2D systems} \label{AppNG}
Before analyzing near-gate plasmons in 2DM with disk-shaped gate, let us remind the properties of these plasmons in 2D system with massive/massless charge carriers in stripe-shaped gate geometry~\cite{zabolotnykh2019} and, then, compare them. We will use standard quasi-static approximation to describe plasmons, without taking into account retardation effects.

Consider infinite 2D electron system with stripe-shaped ideal (i.e. with infinitely large conductivity) metallic gate situated at the distance $d$ above 2D system. The metal stripe has the width $W$ and an infinite length, we assume that $d \ll W$. We assume that the conductivity of 2DM obeys the Drude model $\sigma(\omega)=iD/\pi \omega$ with Drude weight $D=\pi n_se^2/m^*$ in massive systems and $D=e^2v\sqrt{\pi n_s}/\hbar$, where $v$ is the speed of Dirac fermions, in graphene. Note that for the case of graphene the condition $E_F \gg \hbar \omega$ should be satisfied.
Under these assumptions (using the same analytical procedure as in Ref.~\cite{zabolotnykh2019}) the fundamental near-gate plasmon mode at the long wavelength limit $|k_{||}W|\ll 1$, where $k_{||}$ is the wave vector along the stripe, has the dispersion law as follows:
\begin{equation}
\label{Eq:disp_NG}
	\omega(k_{||})=\sqrt{\frac{8 D d}{\epsilon_{\perp}}\frac{|k_{||}|}{W}},
\end{equation}
where $\epsilon_{\perp}$ is the out-of-plane dielectric permittivity between 2DM and the gate. Note that, due to different dependence of $D$ on concentration $n_s$, near-gate plasmon in graphene has typical $n_s^{1/4}$-dependence of frequency on $n_s$, in contrast to $n_s^{1/2}$-dependence for plasmon in massive 2DM.

Near-gate plasmon~(\ref{Eq:disp_NG}) has square-root dispersion law $\omega\propto \sqrt{k}$ similarly to ungated 2D plasmon, which obeys $\omega_{ung}=\sqrt{2D k/\epsilon}$ for 2DM placed in isotropic dielectric medium with permittivity $\epsilon$. Consequently, the relation of the frequencies of these plasmon modes at the same wave vector $k$ does not depend on $k$ and (for isotropic media with $\epsilon_{\perp}=\epsilon$) equals $2\sqrt{d/W} \ll 1$, i.e. frequency of near-gate plasmon is below that of ungated plasmon. That is why near-gate plasmon can not decay into the continuum of ungated 2D plasmons (the frequency and wave vector along the stripe are fixed).

For the near-gate mode discussed above, profile of charge density across the stripe does not have zeros. 
However, there are also higher energy modes having finite number of zeros of charge density across the stripe~\cite{zabolotnykh2019}. These modes possess non-zero frequencies and finite lifetime at $k_{||}\to0$ as they fall into continuum of ungated 2D plasmons. Nevertheless, at large $k_{||}$ these modes are localized near the stripe as well as the fundamental near-gate mode.

Now let us move to near-gate plasmons in 2DM with disk-shaped gate, as shown in Fig.~\ref{fig:fig1}. Their dispersion law (found as poles of the scattering amplitude~(\ref{Eq:scattering_amplitude})) is defined by Eq.~(\ref{Eq:disk_modes_dispersion}). Eq.~(\ref{Eq:disk_modes_dispersion}) has no solutions with real frequency $\omega$, it has only quasi-stationary solutions with complex-valued $\omega$. However in the formal limit $kR \to 0$, which qualitatively corresponds to the suppression of ungated plasmons excitation, as their wave length $2\pi/k$ tends to infinity, one can use standard series for Hankel functions at small arguments in the right-hand side of Eq.~(\ref{Eq:disk_modes_dispersion}) 
to find the dispersion equations as follows
\begin{equation}
\label{Eq:disl_approx}
    k_d R J'_l(k_dR)+ |l| J_l(k_dR)=0 \quad \text{at}\quad l\neq 0
\end{equation}
and 
\begin{equation}
\label{Eq:disl_approx_zero}
    k_d R \ln\left( \frac{e^C kR }{2} \right) J'_0(k_dR)- J_0(k_dR)=0 \,\, \text{at}\,\, l=0, 
\end{equation}
where the stroke defines the derivative by the argument and $l$ is the orbital number. Eqs.~(\ref{Eq:disl_approx}) and (\ref{Eq:disl_approx_zero}) have solutions with real $\omega$. Note, that in this limit, wave vector of ungated plasmons $k$ vanishes in Eq.~(\ref{Eq:disl_approx}), which qualitatively can be understood as that we distinguish the near-gate plasmon modes with $l \neq 0$ from the continuum of ungated 2D plasmons. 
Also, it should be mentioned that Eq.~(\ref{Eq:disl_approx}) coincides with dispersion equation derived in Ref.~\cite{zabolotnykh_disk_2019} by another approach, namely, solving the integral equation for self-consistent potential of near-gate plasmon modes. 
Taking into account more terms in the expansion of Hankel functions in the right-hand side of Eq.~(\ref{Eq:disk_modes_dispersion}), one can obtain additional terms (with imaginary contributions) in Eqs.~(\ref{Eq:disl_approx}) and (\ref{Eq:disl_approx_zero}).

Essentially, roots of Eqs.~(\ref{Eq:disl_approx}) and (\ref{Eq:disl_approx_zero}) define (approximate) real parts of dimensionless frequencies of the modes $k_dR= Re(\omega)R/V_p$ (here $V_p$ is the velocity of gated plasmons defined after Eq.~(\ref{Eq:k_d_limit})), which are given in Table~\ref{tab:quantization}. The radial number $n$ is the root's number of Eqs.~(\ref{Eq:disl_approx}) and (\ref{Eq:disl_approx_zero}), counted from zero. 

As was mentioned above, we assume that the conductivity of 2DM can be described in the Drude model. Then, dimensionless frequencies given in Table~\ref{tab:quantization} are independent of whether 2DM has massive or massless charge carriers. All the difference is in the definition of velocity of gated plasmons $V_p$ (via the Drude weight $D$), which is given after Eq.~(\ref{Eq:k_d_limit}). Note once again that the main characteristic feature of near-gate plasmons in massless 2DM is their $n_s^{1/4}$-dependence of frequency on concentration $n_s$.

\end{document}